\documentclass[12pt,preprint]{aastex}

\newcommand{\Msun}{~M_\odot}
\newcommand{\msun}{M_\odot}

\newcommand{\kms}{\rm ~km~s^{-1}}
\newcommand{\ergs}{\rm ~erg~s^{-1}}

\newcommand{\ml}{~\Msun ~\rm yr^{-1}}

\begin{document}

\title{RADIO AND X-RAY EMISSION AS PROBES
OF TYPE IIP SUPERNOVAE AND RED SUPERGIANT MASS LOSS}
\author{
Roger A. Chevalier\altaffilmark{1},
Claes Fransson\altaffilmark{2},
Tanja K. Nymark\altaffilmark{2},
}
\altaffiltext{1}{Department of Astronomy, University of Virginia, P.O. Box 3818, 
Charlottesville, VA 22903; rac5x@virginia.edu}
\altaffiltext{2}{Department of Astronomy, Stockholm University, AlbaNova, 
SE--106~91 Stockholm, Sweden}


\begin{abstract}
Type IIP (plateau) supernovae are thought to come from stars with initial
mass $\sim 8-25\Msun$ that end their lives as red supergiants.
The expected stellar end points can be found from evolutionary calculations
and the corresponding mass loss properties at this point can be estimated
from typical values for Galactic stars.
The mass loss densities of observed supernovae can be estimated from observations
of the thermal X-ray and radio synchrotron emission that result from the
interaction of the supernova with the surrounding wind.
Type IIP supernovae are expected to have energy-conserving interaction during
typical times of observation. 
Because Type IIP supernovae have an extended period of high optical luminosity,
Compton cooling can affect the radio emitting electrons, giving rise to
a relatively flat radio light curve in the optically thin regime.
Alternatively, a high efficiency of magnetic field production results in
synchrotron cooling of the radio emitting electrons.
Both the X-ray and radio luminosities are
sensitive to the mass loss and initial masses of the progenitor stars,
although the turn-on of radio emission is probably the best estimator
of circumstellar density.
Both the mass loss density and the variation of density with stellar mass
are consistent with expectations for the progenitor
stars deduced from direct observations of recent supernovae.
Current observations are consistent with mass being the
only parameter; observations of a supernova in a metal poor region might
show how the mass loss depends on metallicity.

\end{abstract}


\section{INTRODUCTION}

SNe IIP (Type IIP supernovae) may be relatively straightforward
to interpret in terms of stellar evolution models.
They are characterized by a long, plateau light curve that
can be attributed to the H envelope of the progenitor star.
The progenitor stars have not undergone strong mass loss and
are consistent with $\sim 8-25\Msun$ stars that have
evolved as single stars \citep[e.g.,][]{Heg03}.
Stellar evolution models, models of supernova light curves,
and observations of supernova progenitors all suggest that
these stars explode as red supergiants.

Although only a fraction of the stellar mass is lost, red supergiants
are known to have slow winds that provide a target for the expanding
supernova.
The properties of these winds, based on empirical fits to stellar
observations, are an important input to stellar evolution models.
The supernova interaction with the wind generates a hot, shocked
region that can be observed by X-ray emission and by radio synchrotron
emission from relativistic electrons \citep{Wei86,CF03}.
SNe IIP provide an excellent place to study the interaction because
asymmetries may not be significant and rapid evolution is not
expected before the supernova.
The main parameters determining the mass loss and explosion properties
are expected to be the mass and metallicity of the progenitor star,
although rotation could also play a role.

The expected mass loss rates and hydrodynamic interaction  are described in \S~2.
The resulting radio and X-ray emission properties are discussed in
the context of observations of SNe IIP in \S~3 and \S~4, respectively.
The relation of progenitor mass, deduced from
the recent identifications of the stellar progenitors
of SNe IIP, to mass loss is treated in \S~5.

\section{HYDRODYNAMIC INTERACTION}

Calculations of massive star evolution typically include
mass loss by using a parametrized
fit to observed stellar mass loss rates.
A number of evolutionary calculations \citep[e.g.,][]{Sch92,Heg03}
use the parametrizations of \cite{dNv88} or
\cite{Nd90}.
An older parametrization specifically for cool stars is that
of \cite{R77}.
Fig. \ref{massloss} shows the values of $\dot M$ at the time of the explosion using
the evolutionary tracks of \cite{Sch92}, including the
values given by Schaller et al. based on \cite{dNv88}
and values derived from the expression of \cite{R77}.
The uncertainty in the mass loss rates is estimated by the authors to
be about a factor of 2, which is comparable to the difference between
the two parametrizations.
More recent evolutionary tracks, like those given in \cite{Sma04},
are in close agreement with those given by Schaller et al. (1992).
Although the exact mass limits over which SNe IIP occur are not known,
it can be seen that $\dot M$ varies over about an order of magnitude.

In these models, the progenitor star does not enter a `superwind' phase
at the end of its life.
\cite{ET04} find that  the stellar luminosity can rise at the second dredge-up at the
end of a star's life, presumably giving a rise in the mass loss rate,
perhaps driven by thermal pulses.
The upper initial mass limit for which the dredge-up occurs depends
on convective overshoot: $8\Msun$ limit for overshooting and $10\Msun$
limit for no overshooting.
If much of the envelope is lost, the supernova may not be of Type IIP.
\cite{Chu97} has discussed how supernovae with dense wind interaction
may have progenitors near the low mass limit.

The above results are for solar metallicity, $Z\approx 0.02$, where $Z$ is
the heavy element mass fraction.
Evolutionary calculations typically assume that $\dot M\propto Z^{0.5}$
\citep{Sch92,Heg03},
although this dependence does not have a sound basis in either theory
or observations of cool stars.

Another possible parameter for mass loss is stellar rotation.
\cite{MM00} find a significant dependence of
the mass loss rates as a function of rotation. If we assume that most of
the He-burning takes place in the red supergiant phase (as is true for the
rotating models although the nonrotating low mass models spend some
time as a blue supergiant) we find
$\dot M= (0.84 - 1.6)\times 10^{-6}\ml$ for the $15
\Msun$
models and $(3.0 - 6.2)\times 10^{-6}\ml$ for  the $20 \Msun$ models
for non-rotating and $300 \kms$ rotating models, respectively.
Meynet \& Maeder argue
that the $300 \kms$ models are the more realistic. The values for the
$300 \kms$ rotation models agree with the numbers in Fig. \ref{massloss}, while the
non-rotating values are a factor of two lower.

The important
mass loss parameter for supernova interaction is the wind density, $\rho_w$,
determined by the parameter $A$, where  $\rho_w=Ar^{-2}=\dot M/4\pi v_wr^2$
and $v_w$ is the wind velocity.
The wind velocity for a red supergiant is typically $10-15\kms$.
There is not a well-developed theory for the winds from red supergiants, but
general considerations of winds typically have $v_w$ proportional to
escape velocity, or $v_w\propto M^{1/2}/R^{1/2}$, where $R$ is the
stellar radius.
For the range of cases shown in Fig. \ref{massloss}, the resulting variation in $v_w$
is $<10$\% about a mean value, with the $7\Msun$ star having a high
value and the $20\Msun$ star having a low value.
The variation in wind density is thus only slightly larger than that in
$\dot M$ alone.

For the structure of the supernova, we use the results of \cite{MM99}
for the explosion of a red supergiant.
These models assume that the outer envelope can be approximated
as an $n=3/2$ polytrope, which is accurate if the convection is efficient.
The results for the exploded outer structure depend on $q$, the
fraction of the total progenitor mass that is inside the H envelope.
Since we are mainly concerned with stars that have not undergone
extensive mass loss, we take  $q=0.3$.
We have considered the interaction of the harmonic mean density
profile of Matzner \& McKee with a stellar wind typical of that
expected around a SN IIP and found that the density over the
time of interest ($\sim 1$ yr) is well approximated by the outer
power law density region.
The density in this region can be expressed as $\rho_{SN}\propto
t^{-3}(r/t)^{-n}$ with $n=11.73$.
For this power law profile, the outer shock wave expands as
$R\propto t^{(n-3)/(n-2)}\propto t^{0.90}$.
If the wind interaction is considered in the thin shell approximation
\citep{C82}, the shell velocity is given by
\begin{equation}
V_s=1.6\times 10^4 \left(\dot M_{-6}\over v_{w1}\right)^{-0.10}
E_{51}^{0.45}  \left(M_{ej}\over 10\Msun\right)^{-0.345}
\left(t\over 10{\rm~day}\right)^{-0.10}\kms,
\label{vshock}
\end{equation}
where $\dot M_{-6}$ is the wind mass loss rate in units of
$10^{-6}\ml$, $v_{w1}$ is the wind velocity in units of $10\kms$,
$E_{51}$ is the explosion energy in units of $10^{51}$ ergs,
and $M_{ej}$ is the ejecta mass.
The temperature of the gas at the reverse shock front is
$T_r=1.2 V_{s4}^2$ keV, where $V_{s4}$ is the shell velocity in
units of $10^4\kms$.
The model velocities can be compared to observed velocities in SNe IIP
deduced from absorption lines, which give a minimum value for the
free expansion velocity at the interaction point.
\cite{Elm03} find  a maximum velocity of $\sim
16,000 \kms$ for the blue wing of H-alpha in SN 1999em
at an age of 9 days. 
This velocity is difficult to
determine accurately because of blending with other lines; however, the better
defined absorption minimum is at $12,500 \kms$, lending support to the
above maximum velocity.

Another consideration for the maximum ejecta velocity is that it
may be limited by radiation losses at the time of shock breakout,
independent of the circumstellar interaction.
For a red supergiant with radius $500~R_{\odot}$ and standard supernova
parameters, equation (32) of \cite{MM99} gives a
maximum velocity of $\sim 13,000\kms$, indicating that this could
be a limiting factor.

In the interaction model, the cooling time for gas at the reverse shock is equal
to the age at an age 
\begin{equation}
t_c\approx 20\dot M_{-6} v_{w1}^{-1}V_{s 4}^{-5.2} {\rm days.}
\end{equation}
Substitution of equation (\ref{vshock}) for $V_{s 4}$ yields
\begin{equation}
t_c=0.26 \left(\dot M_{-6}\over v_{w1}\right)^{3.2}
E_{51}^{-4.9}  \left(M_{ej}\over 10\Msun\right)^{3.7} {\rm days,}
\label{eqcool}
\end{equation}
 valid for a reverse shock temperature $\la 2\times 10^7$ K, or for
$n=11.73$ and $V_s \la 12,000 \kms$ \citep{CF03}.  
Considering the dependence of $\dot M$ on mass (Fig. \ref{massloss}), it can
be seen that stars with $M\ga 20\Msun$ have an extended radiative
period.
However, most SNe IIP are expected to have progenitors of lower mass
and to have shocks that are
energy conserving except at very early times,
before observations typically take place.
The presence of radiative cooling would be expected to give a boxy
H$\alpha$ profile at late times, as seen in
SN 1993J \citep{Fil94}; such a profile has never been
reported in a SN IIP.

\section{RADIO EMISSION}
\label{sec_radio}

Radio emission observed from supernovae can be attributed to synchrotron
emission from the interaction region between the supernova and the
surrounding wind \citep{C82}.
The mechanisms for particle acceleration and magnetic field amplification
are not sufficiently well understood to predict the radio luminosity, $L_r$.
However, with the assumptions that the particle and magnetic energy
densities are a constant fraction of the total postshock energy
density, we have
$L_r\propto (\dot M/v_w)^{1.4-2}$ for typical parameters,
when the emission is optically thin and at a fixed age \citep{C82}.
This expression assumes similar supernova properties.

At early times, the radio emission is absorbed; the most likely
mechanisms for SNe IIP are FFA (free-free absorption) by ionized gas
outside the forward shock front and SSA (synchrotron self-absorption).
Optical depth unity to free-free absorption is attained at an age
\begin{equation}
t_{ff}\approx 6
\left(\dot M_{-6}\over v_{w1}\right)^{2/3} T_{cs5}^{-1/2}V_{s 4}^{-1}
\left(\nu\over 8.46{\rm~GHz}\right)^{-2/3}{\rm~days},
\end{equation}
where $T_{cs5}$ is the circumstellar temperature in units of $10^5$
K. The value of $T_{cs}$ is uncertain. Calculations, mainly directed
to Type IIL supernovae, by \cite{LF88} find that for $\dot M_{-6}/
v_{w1}=3$ the temperature at the time of optical depth unity was $\sim
3\times10^4$ K, while for $\dot M_{-6}/ v_{w1}=10$ it was found that
$T_{cs} \sim 1\times10^5$ K. These calculations used a simplified
model for the shock breakout radiation, and may have underestimated
the temperature somewhat. In the following we will use $T_{cs} =
1\times10^5$ K as a reference value, but we emphasize that 
this parameter is uncertain.

A number of recent, well-observed SNe IIP are listed in Table 1.
The distances to SN 1999em and SN 2004dj are based on Cepheids; for
SN 1999em, this distance is 50\% larger than the distance determined
by the EPM (expanding photosphere method) \citep{Leo03}.
The EPM distance to SN 1999gi is used; in this case, the
Cepheid distances to nearby galaxies are only slightly larger
\citep{Leo02}.
The distance to SN 2002hh and SN 2004et (both in NGC 6946) is based on
an average of 3 methods \citep{Li05a}.

The positions of the SNe IIP, as well as other supernovae, in a
peak radio luminosity - time of peak diagram are shown in Fig. \ref{snae}, which
is an update of Fig. 4 of \cite{C98}.
The times of peak of the SNe IIP were determined from plausible
fits to the available data.
It can be seen that the SNe IIP cluster in a small region of
the diagram.
The constant velocity lines in the diagram are based on the assumptions that SSA
is the dominant absorption mechanism and that there
is energy equipartition between particles and fields (although the results
depend only weakly on this assumption).
If the velocity deduced in this way is compatible with the maximum velocities
in the supernova, SSA is a plausible absorption mechanism.
If the velocity is lower than the apparent maximum velocity, another mechanism,
such as FFA, must be operating.
However, the model assumes that $p=3$ and that there is no cooling of radio
emitting electrons; as shown below, these assumptions are probably not
accurate.
The position of the SNe IIP, somewhat below the $10,000\kms$ line, is close
to the velocities expected in SNe IIP, indicating that SSA may play a role
but other mechanisms, such as FFA, may also be important.

Depending on the parameters characterizing the supernova ejecta, the
circumstellar medium, 
the magnetic field and the relativistic electrons, cooling of the
relativistic electrons may be important. As we show below, this has
important consequences for both the light curve and the spectrum. 

We first consider synchrotron cooling. Assuming that the magnetic
field energy density is a fixed fraction $\epsilon_B$ of the 
the thermal energy behind the circumstellar shock, the
ratio of the synchrotron time scale to the expansion time scale is
given
\begin{equation}
{t_{synch} \over t}\approx 2.0
\left({\epsilon_B \over 0.1}\right)^{-3/4}
\left({\dot M_{-6}\over v_{w1}}\right)^{-3/4} 
\left({\nu\over 10{\rm~GHz}}\right)^{-1/2} 
\left({t\over 10{\rm~days}}\right)^{1/2}.
\label{tsynch}
\end{equation}
Therefore, in the restricted range $\dot M_{-6}/v_{w1} \ga 4$ and
$\epsilon_B \ga 0.1$ synchrotron losses may be important.

Inverse Compton cooling of the relativistic electrons by the strong
photon flux from the ejecta can be important especially during  the
early phases. The ratio of the Compton cooling time scale and
expansion time scale is
\begin{equation}
{t_{comp} \over t}\approx 0.18
\left({L_{\rm bol} \over 2 \times 10^{42} \ergs}\right)^{-1}
\left({\epsilon_B \over 0.1}\right)^{1/4}
\left({\dot M_{-6}\over v_{w1}}\right)^{1/4} V_{s 4}^{2}
\left({\nu\over 10{\rm~GHz}}\right)^{-1/2} 
\left({t\over 10{\rm~days}}\right)^{1/2},
\label{tcomp}
\end{equation}
where $L_{\rm bol}$ is the bolometric luminosity of the
supernova. Depending on $L_{\rm bol}, \dot M$, and the highly uncertain
value of $\epsilon_B$, inverse Compton cooling may or may not be
important. In general, low values of $\epsilon_B$ and $\dot M$ favor
inverse Compton cooling.

There are few well determined bolometric light curves of Type IIP
supernovae. Observations of the normal Type IIP SN 1992H show that
$L_{\rm bol}$ decreased from $\sim 3\times 10^{42} \ergs$ to $\sim
1.3\times 10^{42} \ergs$ during the first $\sim 50$ days, and then
stayed constant up to $\sim 100$ days, when it dropped
\citep{Clocc96}. There is, however, a considerable range in
luminosities within the IIP class. \cite{H03} finds that the absolute
magnitude during the plateau phase varies between $M_V=-15.37$ and
$M_V= -18.57$, with $M_V=-17.41$ for SN 1992H.  From modeling of  
SNe IIP, \cite{East94} find that at day one  $L_{\rm bol} \sim 2.5\times
10^{42} \ergs$, and during the plateau phase $L_{\rm bol} \sim (1-2)\times
10^{42} \ergs$, in good agreement with SN 1992H.  \cite{Chi03}
find a similar luminosity for SNe IIP during the first 50 days. The
energy and envelope mass of the models are, however, adjusted to fit
the observations, and are not calculated from first principles.
Adopting a luminosity of $L_{\rm bol} \sim 1\times 10^{42} \ergs$, we
   find that for $\epsilon_B \la 0.1$ inverse Compton cooling
is likely to be important during most of the plateau phase, except for
very high mass loss rates or shock velocities.

Because $t_{comp} \propto \epsilon_B^{1/4}$ and $t_{synch} \propto
\epsilon_B^{-3/4}$ the Compton timescale is likely to be short in the
cases where the synchrotron timescale is long, and vice versa.
In Figure \ref{figcomp} we show how the cooling timescales as function of
the parameter $\epsilon_B A_*\equiv \epsilon_B \dot M_{-5}/v_{w1}$. Because both the
synchrotron and Compton cooling timescales are proportional to
$(t/\nu)^{0.5}$ we plot the scaled cooling timescale $t_{cool}
(t_{10}/\nu_{10})^{0.5}/t$, where $t_{10}=t/10 $ days and
$\nu_{10}=\nu/10$ GHz. The remaining parameter is then $l_{42} =
(L_{\rm bol}/10^{42} {\ergs})(V_s/15,000 \kms)^{-2}$.

As an example, for $\dot M_{-5}/v_{w1}=1$, $\epsilon_B=0.1$, $\nu=10$
GHz, $l_{42}=1$ and $t=10$ days we get $t_{\rm comp}/t \approx 1$ and
$t_{\rm synch}/t \approx 0.3$. For the same parameters, but with
$\epsilon_B=10^{-3}$, we instead obtain $t_{\rm comp}/t \approx 0.3$ and
$t_{\rm synch}/t \approx 10$. In both cases the electrons are cooling,
but in the former case by synchrotron cooling and in the latter by
Compton cooling. 

The fact that cooling is likely to be
important in many cases for the relativistic electrons means that the spectral index
of the radio emission (flux $\propto \nu^{-\alpha}$) 
is expected to be steep with $\alpha \approx
1.0$, if the particles are injected with an energy spectral index $p\approx 2$.
As cooling becomes less important at late times, the spectrum should
flatten toward $\alpha\approx 0.5$. 
This is one of the important diagnostics of cooling.

To show the possible range of radio light curves, we have calculated
radio light curves for four different cases, two where cooling is
important, one `minimal cooling' case, and one purely adiabatic
case. All calculations have $p=2.2$, $\dot M_{-6}/v_{w1} =5$,
$n=11.73$ and $V_s=15,000 \kms$ at 10 days. For the bolometric light
curve we take that of SN 1992H and scale it by a factor $f_{\rm
bol}$. For each value of $f_{\rm bol}$ we vary $\epsilon_B$ and
$\epsilon_r$ to give a peak flux of 2 mJy at 4.9 GHz at 5 Mpc, typical
of the observed Type IIP light curves.  Figure \ref{figcomp} and
equations (\ref{tsynch}) and (\ref{tcomp}) show  that one can
have two qualitatively different cases of cooling. For high values of
$\epsilon_r$ (and therefore small values of $\epsilon_B$) Compton
cooling dominates. Conversely, large values of $\epsilon_B$ imply
synchrotron cooling. We therefore calculate two models typical of
these cases, one, model C, with $f_{\rm bol}=1$, $\epsilon_B=0.0013$
and $\epsilon_r=0.05$, and one, model S, with $f_{\rm bol}=1$,
$\epsilon_B=0.15$ and $\epsilon_r=0.001$.

To find parameters relevant for a realistic non-cooling case is  
non-trivial. To minimize inverse Compton losses, and still be within the
range of reasonable bolometric luminosities, we take $f_{\rm
bol}=0.05$ and $V_s=20,000 \kms$. During the plateau phase this
corresponds to $L_{\rm bol} \sim 1\times 10^{41} \ergs$, giving
$l_{42} = 5.6\times 10^{-2}$. From Figure \ref{figcomp} we see that
minimum cooling then occurs for $\epsilon_B \dot M_{-5}/v_{w1}
\approx 2\times 10^{-3}$. With $\dot M_{-5}/v_{w1}=1$, to reproduce
the approximate turn-on time we require $\epsilon_r \approx 1\times
10^{-2}$ to give the approximate flux level. We refer to this as model
M. Finally for comparison, we show one model, model A, where we have artificially
put both synchrotron and inverse Compton cooling  to
zero. In Figure \ref{figlc} we show the resulting light curves for 5.0,
8.46, and 22.5 GHz.

The most interesting aspect of these light curves is the variation in
the optically thin decline rate. While the flux in the adiabatic
model, model A, falls off as $F_\nu \propto
t^{-(p+5-6m)/2}=t^{-0.8}$, the others have a flatter
evolution. Here $m=(n-3)/(n-2)=0.9$ for $n=12$. This is especially
true for Model C, which is dominated by Compton cooling of the
electrons. Because the Compton cooling rate depends on the bolometric
luminosity of the supernova, the radio light curve reflects the shape
of the bolometric, primarily optical, light curve.  
In the limit that $t_{comp} \ll t$ the stationary solution
of the kinetic equation for the relativistic electron distribution
\citep{FB98}, together with the expression for the synchrotron flux,
shows that
\begin{equation}
F_\nu \propto
L_{\rm bol}(t)^{-1} t^{m-p/2}
\nu^{-p/2}.
\label{fcomp}
\end{equation}
This shows that the synchrotron flux in this limit is inversely
correlated with the bolometric luminosity. Further, for $m\approx 1$
and $p\approx 2$ one obtains a flat light curve in the
plateau phase of the bolometric light curve.  Note also the dip in the
numerical light curves at $\sim 100$ days, seen especially at the highest
frequencies. This coincides roughly with the time when the electrons
become adiabatic, which at 5.0 GHz occurs at $\sim 80$ days for these parameters.

In the synchrotron cooling dominated model S, the cooling has in
contrast a smooth evolution, $t_s\propto t^{1/2}$. Consequently, the
radio light curve does not show any features like the Compton
dominated case. Finally, we note that the
minimum cooling case, model M, has a flatter decline than the purely
adiabatic case, showing that cooling even for this case cannot be
neglected.

The only SNe IIP with sufficient radio data available to compare to
our models are SN 2004et and SN 2004dj.
In other cases, there are some data on the radio rise, which can
be used to estimate the circumstellar density from FFA.

\subsection{SN 2004et}

The data on SN 2004et, shown in Figure \ref{sn2004et_lc_fit}, represent
the most complete time coverage of any SN IIP. The data are from
\cite{Sto04c} before day 20, with additional data at 4.99 GHz after
day 20 from \cite{Bes04} and \cite{Arg05}. Some additional upper
limits at early epochs exist for the 4.99 GHz observations, but these
do not add any strong constraints.  Because of the good time
coverage we have done a more detailed analysis of this supernova than
for the other Type IIPs. 

The explosion date of SN 2004et is well constrained to be 22.0
September, 2004, within one day \citep{Li05a}.  The light curve of SN
2004et showed that it was a Type IIP supernova, although it showed
some differences from typical Type IIP events \citep{Li05a}. For the
simulations below we use the bolometric light curve of SN 1992H,
scaled with $f_{\rm bol}=0.5$.  From the spectra of \citet{Li05a} we
find a maximum velocity of $\sim 14,200 \kms$ for the red wing of the
H$\alpha$ line at days 9 and 20.  Because of occultation and other effects
this is only a lower limit, and in the following we will use an
expansion velocity of $15,000 \kms$ at 10 days.  The relative turn-on
times between the different frequencies depend on the value of $n$,
and we here take $n=10$. A smaller value may give a better agreement
with the observed turn-on times, but would also result in steeper
decays in the optically thin parts of the light curves. All models
have an injected electron spectrum with $p=2.2$. Cooling
considerably steepens the integrated spectrum.

As discussed above, we vary the value of $\epsilon_B$, and
consequently $\epsilon_r$, to cover both a Compton dominated and a
synchrotron dominated case. In Figure \ref{sn2004et_lc_fit} we 
show two sets of light curve fits with either large or small values of
$\epsilon_B$ and $\epsilon_r$, $\epsilon_B=0.2$ and
$\epsilon_r=0.001$, and $\epsilon_B=0.0013$ and
$\epsilon_r=0.04$, respectively.  The value of $\dot M_{-6}/v_{w1}$ is
determined by the turn-on time, but, as pointed out before, also
depends on the uncertain temperature of the CSM. In the first case we
obtain $\dot M_{-6}/v_{w1}= 9~  T_{cs5}^{3/4}$, and in the second $\dot
M_{-6}/v_{w1}= 10 ~ T_{cs5}^{3/4}$. 

The two cases result in fairly similar light curves, both strongly
affected by cooling during the first $\sim 100$ days. The main
difference is that for $\la 20$ days the Compton dominated case shows
a lower peak flux for the higher frequencies compared to the
lower. This is caused by the decreasing Compton cooling rate with
time, which affects the higher frequencies, which first become
optically thin and also have a higher Lorentz factor of the
electrons. In addition, as we have already found, at later epochs the
light curves are flatter than for the synchrotron dominated case. The
considerable error bars of the observations, especially at early time,
make it difficult to favor one of these cases over the
other. Because of the difference in decay slope it is of
great interest to pursue these observations to as late times as
possible to check these alternatives. The frequency dependent dip in
the light curve at $\sim 100$ days in the Compton case would be an
interesting signature. However, effects
connected with  a gas density gradient different from a
simple $\rho \propto r^{-2}$ wind may also influence the late time
behavior and high observational accuracy would be required to
detect the dip.

In the $\epsilon_B=0.0013$,
$\epsilon_r=0.04$ model, the optical depths to synchrotron self-absorption
and free-free absorption are similar, while in the $\epsilon_B=0.2$,
$\epsilon_r=0.001$ model the SSA optical depth is larger than 1 when
it becomes thin to FFA. Therefore, especially in the latter model the
light curve and spectrum are affected by SSA, giving a slower, more gradual
rise compared to pure FFA.

\subsection{SN 2004dj}

A spectrum of SN 2004dj at the beginning of August 2004
indicated a normal Type IIP supernova (very similar to
SN 1999em) with an age of about 3 weeks \citep{Pat04}.
We take 1 August, 2004 to correspond to an age of 21 days.
The radio data in this case include a point at 8.4 GHz \citep{Sto04a},
 two at
1.4 GHz \citep{CR04}, and more extensive coverage at 4.99 GHz \citep{Bes05}.

As in the case of SN 2004et, there is no single unique combination of
$\epsilon_B$ and $\epsilon_r$ that fits the observations. In
Fig. \ref{sn2004dj_lc_fit} we show two examples, one with small
$\epsilon_B$ and large $\epsilon_r$, and one for which the opposite is
the case. The mass loss rate parameter is rather well
constrained by  the rising 1.4 GHz points, due to decreasing
absorption.  The observed sharp rise between days 33 and 43 is
suggestive of FFA, although SSA cannot be ruled out. In fact, the
$\epsilon_B=0.01$, $\epsilon_r=0.0009$ model is dominated by SSA,
while FFA is dominant for the $\epsilon_B=0.00075$, $\epsilon_r=0.01$
model. We find that $\dot M_{-6}/v_{w1}= 2-3~  T_{cs5}^{3/4}$ for the two
models. 

\subsection{SN 2002hh}

Radio data on SN 2002hh  are available from \cite{Sto04b}
and \cite{CRB04}
for data at 1.4 GHz.
The supernova was discovered on 31.1 October, 2002 and a spectrum
on 2 November, 2002 showed that it was a very young, highly
reddened Type II supernova \citep{FFS02}.
We take the age at the time of discovery to be 2 days.
There are no published optical light curve observations, as yet, that
identify SN 2002hh as a plateau type.
However, optical spectra of the  supernova in the nebular phase
show that it has slow moving H \citep{Mat04}; the mixing of H to low velocities
is expected to require a massive H envelope compared to the core,
as occurs in a SN IIP.
Also, its radio properties are
similar to other SNe IIP.

The radio rise suggests optical depth unity to FFA at 1.4 GHz on day
62.  Table 2 lists the resulting value of $\dot
M_{-6}v_{w1}^{-1}T_{cs5}^{-3/4}$.  This quantity is $\propto
V_{s4}^{3/2}$; the value $V_{s4}=1.5$ at 10 days has been taken and
allowed to evolve $V_{s4}\propto t^{-0.1}$.  At the relatively low
circumstellar density for SN 2002hh and other SNe IIP, Compton heating
to $\sim 5\times10^4 - 10^5$ K occurs and recombination is not
important   \citep[see above,][]{LF88}. 

A different picture for the circumstellar medium of SN 2002hh comes from
its infrared detection 
which \cite{Bar05} find is due to dust in
circumstellar mass loss $\sim 5\times 10^{17}$ cm from the supernova
based on the infrared spectrum and luminosity.
A similar radius for the dust can be obtained by considering that
the duration of the infrared emission should be comparable to the light
travel time across the emitting region and the total radiated energy
should be approximately that radiated optically by the supernova.
\cite{Bar05} estimate the dust mass to be $0.10-0.15\Msun$, corresponding to
a gas mass of $10-15\Msun$, and
 attribute a significant part of the $A_V\approx 5$
extinction toward the supernova to this dust.
Supernova dust echoes were first inferred around SN 1979C and SN 1980K
for which \cite{D83} found circumstellar optical depths of 0.3
and 0.03 respectively.
Scaling with mass loss density and using our value of $\dot M/v_w$
leads to an optical depth $\la 0.01$
for SN 2002hh, which would not contribute a significant extinction.
An estimate of the mass loss rate from the mass and size of the dust
emitting region also leads to a rate $\sim 10^2$ times larger than
that estimated from the radio emission.

In principle, there is no contradiction between these observations;
the radio emission samples a region inside of the dust emitting region.
There may be a rise in the radio emission in $15-20$ years.
However, the evidence for the loss of $10-15\Msun$ of gas is difficult
to reconcile with the slow moving H in the supernova, which would seem to require a
massive H envelope at the time of the explosion.

\subsection{Other Supernovae}

SN 1999em was a well-observed, normal Type IIP supernova, including
evidence for slow H in the nebular phase \citep{Elm03}.
The radio and X-ray data on it have
already been discussed in terms of circumstellar interaction
by \cite{Poo02}.
The peak in the light curve is suggested by the comparable fluxes at 1.4 and 4.9 GHz
at days $50-70$ \citep[see Fig. 2 of][]{Poo02}.
If not for these, the data could only give an upper limit for the
time of peak flux.
However, the agreement of the results here with those from the other
supernovae suggests that the flux peak was observed.
An estimate of the mass loss rate is given in Table 2.

The low luminosities of the detected SNe IIP show that the published upper
limits that have been set for radio supernovae \citep[e.g.,][]{Wei89,ECB02}
are above the
expected flux levels for supernovae of this type.
However, an upper limit of 0.07 mJy (8.46 GHz)
for SN 2003gd on 17 June, 2003 \citep{S03} is more stringent.
The age of the SN IIP was $\sim  87$ days at that
time \citep{VLF03}.
Assuming models similar to those discussed above, the limit is consistent
with the expected flux of SN 2004dj, but is a factor $\sim 3$ lower
than the expected fluxes of SNe 2002hh and 2004et.
These results suggest an upper limit
$\dot M_{-6}/v_{w1}\la 3$.

\section{X-RAY EMISSION}

We first consider the thermal emission from the shocked region.
The monochromatic luminosity of the circumstellar shock can be
estimated from 
\begin{equation}
\label{eqnLcs}
\frac{dL_{\mathrm{cs}}}{dE} \approx 6.9\times 10^{35}\zeta
T_9^{-0.08} E_{\rm keV}^{-0.23} e^{-0.0116 E_{\rm keV}/T_9} \left(\dot
M_{-6}\over v_{w1}\right)^{2} V_{s4}^{-1}\left (\frac{f}{0.2}\right
)^{-1}\\ \nonumber \\ \nonumber \left(\frac{t}{10\
{\mathrm{days}}}\right )^{-1}\ \ \mathrm{ergs\ s^{-1} ~keV^{-1}}
\end{equation}
\citep{Nym2005}. Here $T_{9}=T_{\mathrm{e}}/10^9$~K, $E_{\rm keV}$ is the photon energy
in keV, $f$ is the relative thickness of the circumstellar shock
region, and $\zeta =[1+2n(\mathrm{He})/n(\mathrm{H})] /
[1+4n(\mathrm{He})/n(\mathrm{H})] \approx 0.86$ for solar
composition.   For $n =7$ we have $f\approx 0.3$, and for $n \ga 12$,
$f\approx 0.22$ \citep{CF94}. The electron temperature,
$T_{\mathrm{e}}$, is uncertain because of the unknown importance of
collisionless heating. If only Coulomb collisions are important
$5\times 10^8 \la T_{\mathrm{e}} \la 1\times 10^9$ K. The luminosity from
equation~(\ref{eqnLcs}) should be added to the flux from the reverse
shock below.

If the reverse shock is in the nonradiative regime and free-free
emission dominates, which is a good approximation for $T \ga 5\times
10^7$ K, the X-ray luminosity at $\sim 1$ keV is given by
\citep{FLC96}
\begin{equation}
\frac{dL_{\mathrm{rev}}}{dE }=2.0\times 10^{35}\zeta (n-3)(n-4)^2 T_8^{-0.024} e^{-0.116/T_8}
\left(\dot M_{-6}\over v_{w1}\right)^2 V_{s4}^{-1}
\left(t\over 10{\rm~days}\right)^{-1} {\rm~ ergs~s^{-1}~keV^{-1}}.
\label{xray}
\end{equation}
 The main variation of X-ray luminosity
with wind density is the $(\dot M_{-6}/ v_{w1})^2$ term. Both equations
(\ref{eqnLcs}) and (\ref{xray}) assume that the ingoing half of the
X-ray emission is absorbed by the ejecta.  In general, 
the emission from the reverse shock dominates in the
keV region, unless the shock is radiative so that absorption by the
cool dense shell is important.

For $T \la 5\times 10^7$ K, line emission becomes important and a more
detailed calculation of the spectrum of the reverse shock is
required. For this we have used the code described in \citet{Nym2005} which
calculates the ionization and emission for a given electron
temperature and density. Because the reverse shock, for the mass loss
rates discussed here, is nonradiative, we assume that the ion
temperature and density are given by the self-similar structure for the
particular value of $n$. The electron temperature is calculated 
assuming Coulomb collisions only. For the reverse shock, this is enough
to bring electrons and ions into equipartition. The ionization is
calculated from the time a given element is shocked to the observed
epoch, using the full time dependent equations. Because only a
fraction of the hydrogen envelope has been lost for the Type IIP
supernovae, we  assume solar composition. 

In Fig. \ref{figxray} we show the   resulting X-ray spectra for
$n=7$ and 12, and $V_s=10,000 \kms$ and $V_s=15,000 \kms$. The mass
loss rate is assumed to be $5\times 10^{-6} \ml$ (for $v_{w} =
10\kms$). Because $dL/dE \propto (\dot M_{-6}/v_{w1})^{2}$ as long as
cooling is not important, the spectra can be scaled to other
mass loss rates, using this relation. The most interesting thing to
note from Fig. \ref{figxray} is the increasing importance of line
emission as the reverse shock temperature decreases. For the models
shown, $T_{\rm rev}=8.5\times 10^7$ K for $n=7$, $V_s=10,000 \kms$;
$T_{\rm rev}=1.9\times 10^8$ K for $n=7$, $V_s=15,000 \kms$; $T_{\rm
rev}=1.7\times 10^7$ K for $n=12$, $V_s=10,000 \kms$; and $T_{\rm
rev}=3.8\times 10^7$ K for $n=12$, $V_s=15,000 \kms$. While 
the line complex at $\sim 1$ keV is especially strong in the $n=12$ models, it
is absent in the $n=7$ models, where the only lines present are
the Fe K lines at $\sim 7$ keV. The feature at $\sim 1$ keV is due to Fe
XXI--XXIV, O VIII, and Ne IX--X, in this order. 

In Fig. \ref{figxray} we have not included any absorption by the
cooling gas behind the reverse shock, the circumstellar medium, or interstellar
gas. The first two of these depend on the mass loss rate assumed. The
total column density behind the reverse shock is 
\begin{equation} 
N_{\rm rev} 
\approx 1.0\times 10^{21} (n-4)  ~\left({\dot M_{-6} \over  
v_{\rm w1} }\right) V_{s4}^{-1} \left({t \over  
10 \rm ~days}\right)^{-1}~\rm cm^{-2}.
\label{eq11b}
\end{equation}
In the case that the gas is cooling (i.e. for $\dot M_{-6} \ga 5$ and
for early epochs, see eq. [\ref{eqcool}]) the emission from the
reverse shock will be absorbed below $\sim 1.2 (N_{\rm rev}/10^{22}
\rm cm^{-2})^{3/8}$ keV. Even if the gas is not cooling and $T_{\rm
rev} \la 2\times 10^7$ K, incompletely ionized atoms like
Mg, Si, and Fe may in principle contribute to absorption {\it above}
$\sim 2$ keV \cite[e.g.,][]{KK84}. However, the opacity at 2 keV is 
 only $\sim 2.5 \times 10^{-24} \ \rm cm^2$ for $\sim 10^7$ K, and
absorption is unlikely to be important for reasonable parameters
unless the gas is cooling.

Both equation (\ref{xray}) and Fig. \ref{figxray} show that the
luminosity is sensitive to $n$.  We found from the models for the
radio emission for SNe IIP that $n\ga 12$ is indicated (\S~3).  However, for the
best observed X-ray SN IIP SN 1999em, \cite{Poo02} suggested $n=7-9$ based
on studies of the optical spectrum and the moderately high temperature
($kT=5$ keV) found from the X-ray spectrum.  The result was $\dot
M_{-6}/v_{w1}=1-2$; a higher value of $n$ would lower the mass loss
density, bringing it further from agreement with the value deduced
from the radio observations (Table 2).

X-rays may also be produced by inverse Compton emission. As we have
seen in \S~\ref{sec_radio}, inverse Compton cooling by the
photospheric photons may be important. These photons are up-scattered
to $\sim 3.1
\gamma^2 (T_{\rm eff}/10^4 \rm K) $ eV, where $\gamma$ is the Lorentz
factor of the relativistic electrons and $T_{\rm eff}$ the effective
temperature of the photospheric emission \citep{FM66}. Using the
expression for the inverse Compton emissivity in \cite{FM66}, one can
find the X-ray luminosity, which for the interesting case $p=3$
(cooling important) takes a simple form,
\begin{equation} 
\frac{dL_{\mathrm{IC}}}{dE } 
\approx 8.8\times 10^{38} \ \epsilon_r \gamma_{\rm min} E_{\rm
  keV}^{-1} ~\left({\dot M_{-6} \over  
v_{\rm w1} }\right) V_{s4} \left(L_{\rm bol}(t) \over 10^{42} \ergs
\right)
 \left({t \over 10 \rm ~days}\right)^{-1}~{\rm~ ergs~s^{-1}~keV^{-1}}.
\label{eq11}
\end{equation}
Here $\gamma_{\rm min}$ is the minimum Lorentz factor of the
relativistic electrons. For other values of $p$, $dL_{\rm IC}/dE
\propto E^{-(p-1)/2}$. The distinguishing property is the power law form,
which is the same as that of the optically thin radio spectrum, and
the correlation of the time evolution with that of the
bolometric luminosity. It is unaffected by any absorption by cool
gas behind the reverse shock. An observation of an inverse Compton
component would allow a determination of the important parameter
$\epsilon_r \gamma_{\rm min}$, and then from the radio observations
$\epsilon_B$. 
The stronger dependence of thermal emission on the circumstellar
density implies that thermal emission is relatively more important at
high $\dot M/v_w$ and inverse Compton at low $\dot M/v_w$. At high
mass loss rates, when reverse shock cooling becomes important,  the
inverse Compton component may again dominate, especially at early epochs.

The observed X-ray luminosities are given in Table 1.
The data are all from {\it Chandra} and refer
to an age of $\sim 30$ days.
SN 2004dj has an X-ray luminosity slightly higher that SN 1999em and
a radio luminosity that is slightly lower (Table 1).
This suggests that they have comparable wind densities.
Both the radio turn-on and the X-ray luminosity of SN 2002hh
indicate that the circumstellar density is somewhat larger in
this case.
The low X-ray luminosity of SN 1999gi suggests a smaller density
in this case (by a factor $\sim 3$), if the supernova properties
are similar to the other cases.

Because of the comparatively low fluxes, the spectral information from
the observed supernovae is limited. A detailed comparison with the
model spectra, or even a discrimination between thermal and
non-thermal spectra, is therefore not possible. Our calculations do,
however, show that deeper observations of nearby SNe IIP can give very
important constraints on both the mass loss and the shock physics.

\section{PROGENITOR MASS AND MASS LOSS}

Progenitor mass estimates are from observations of the region
of the supernova before the supernova occurred.
Studies of this kind have been possible for a number of
nearby SNe IIP (see Table 1).
In the cases of SN 2003gd and SN 2004et \citep{Li05a},
progenitor stars have been tentatively identified and, in the
case of SN 2004dj, a mass was determined from the properties of a
tight cluster at the position of the supernova \citep{Mai05}.
Both SN 2004dj and SN 2004et are estimated to have progenitor masses
$\sim 15\Msun$ for which standard mass loss prescriptions give
$\dot M_{-6}=1.5-3$; the range is increased to $1-7$ when the mass
uncertainty for SN2004et is included.
The mass loss rate deduced from radio observations of SN 2004dj is consistent
with expectations, while that for SN 2004et suggests a mass $\sim 20\Msun$.
However, for SN 2004et \cite{Li05a} find that their tentative progenitor
is a yellow, rather than red, supergiant and may have some similarity to
the peculiar SN 1987A and SN 1993J, although less extreme.
These supernovae had very different radio properties from the normal
SNe IIP (see Fig. \ref{snae}), which have properties close to those of SN 2004et.

\begin{figure}[!hbtp]
\plotone{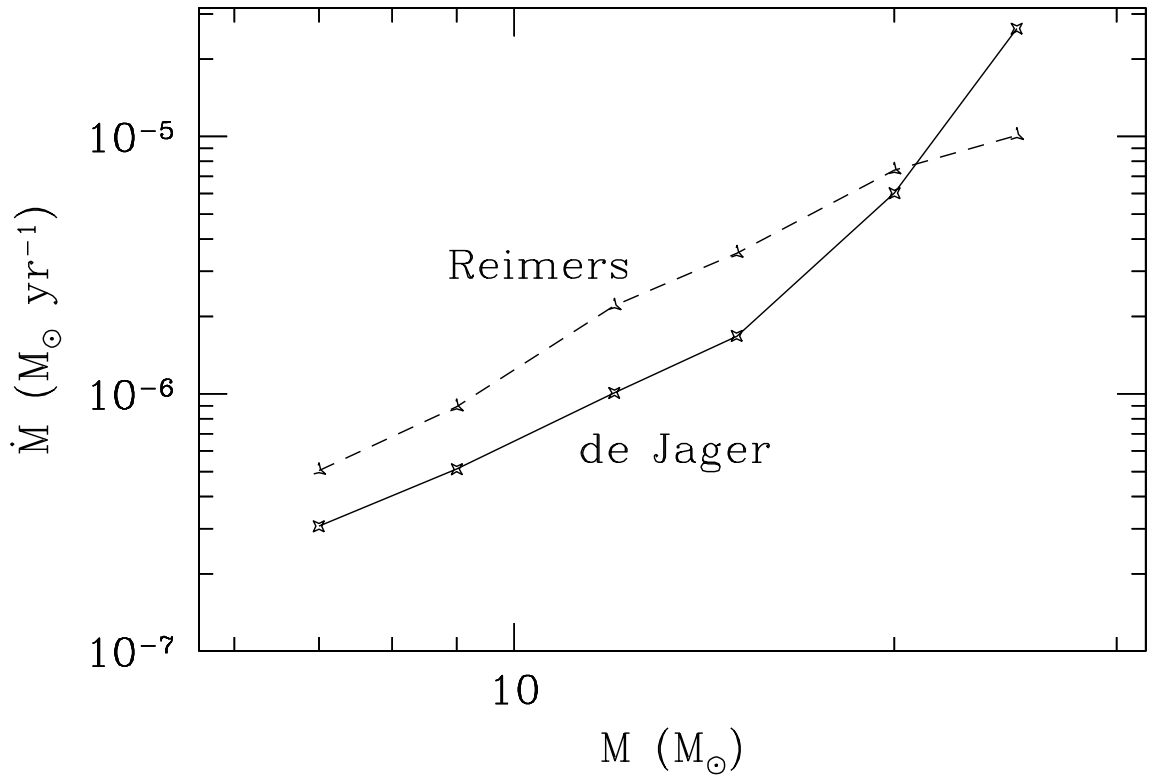}
\figcaption{Mass loss rates from red supergiants at the end of their
lives as a function of initial mass.  The stellar endpoints are from
$M=7, 9, 12, 15, 20, 25\Msun$ evolutionary tracks by Schaller et al. (1992),
and the mass loss rates are based on de Jager et al (1988) ({\it solid})
and Reimers (1977) ({\it dashed}).\label{massloss}
}
\end{figure}

\begin{figure}[!hbtp]
\plotone{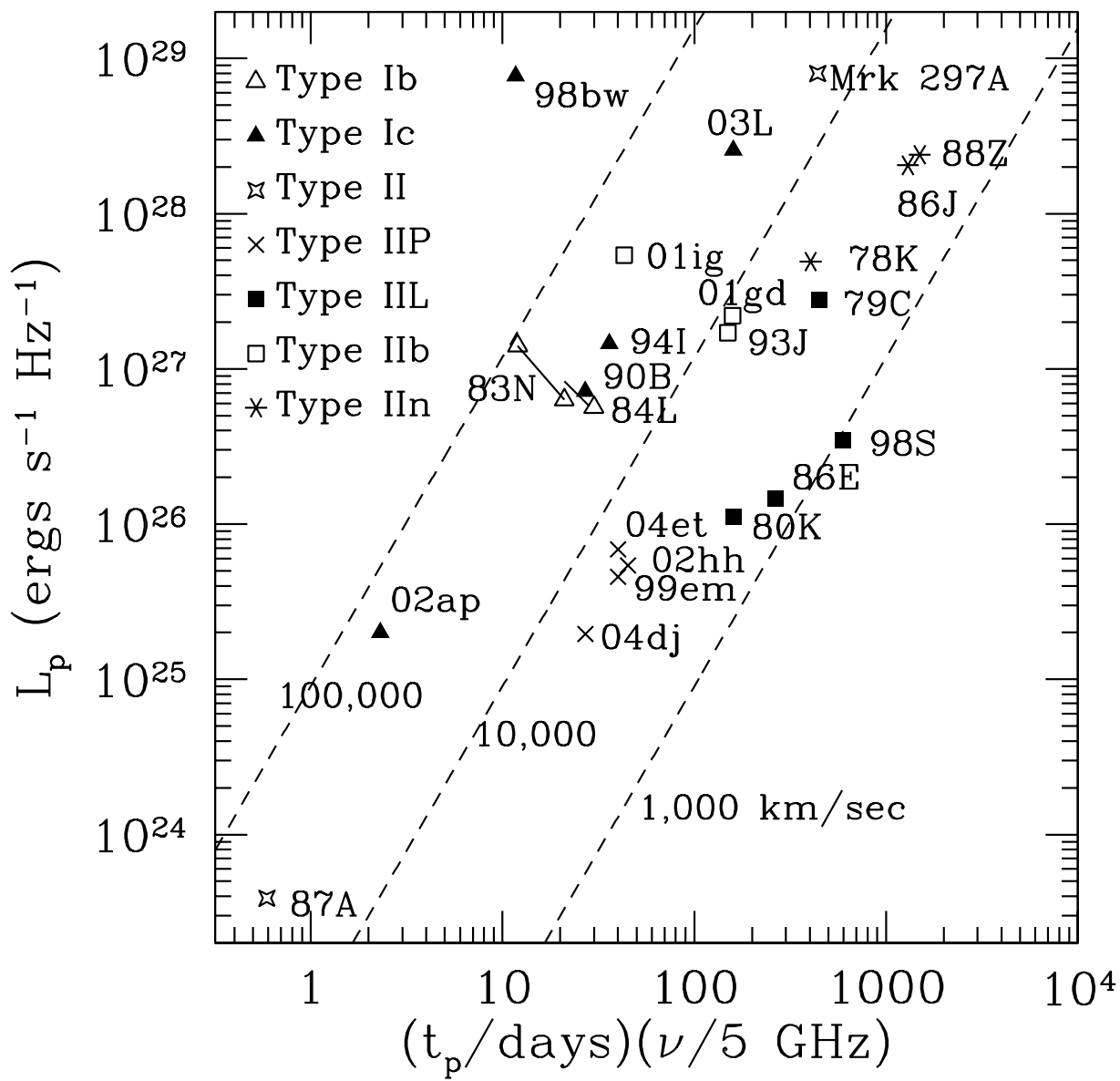}
\figcaption{Peak radio luminosity vs. time of peak for radio supernovae.
The 4 Type IIP supernovae ({\it crosses}) are near the middle of the
plot. 
\label{snae}}
\end{figure}

\begin{figure}[!hbtp]
\includegraphics[scale=.70]{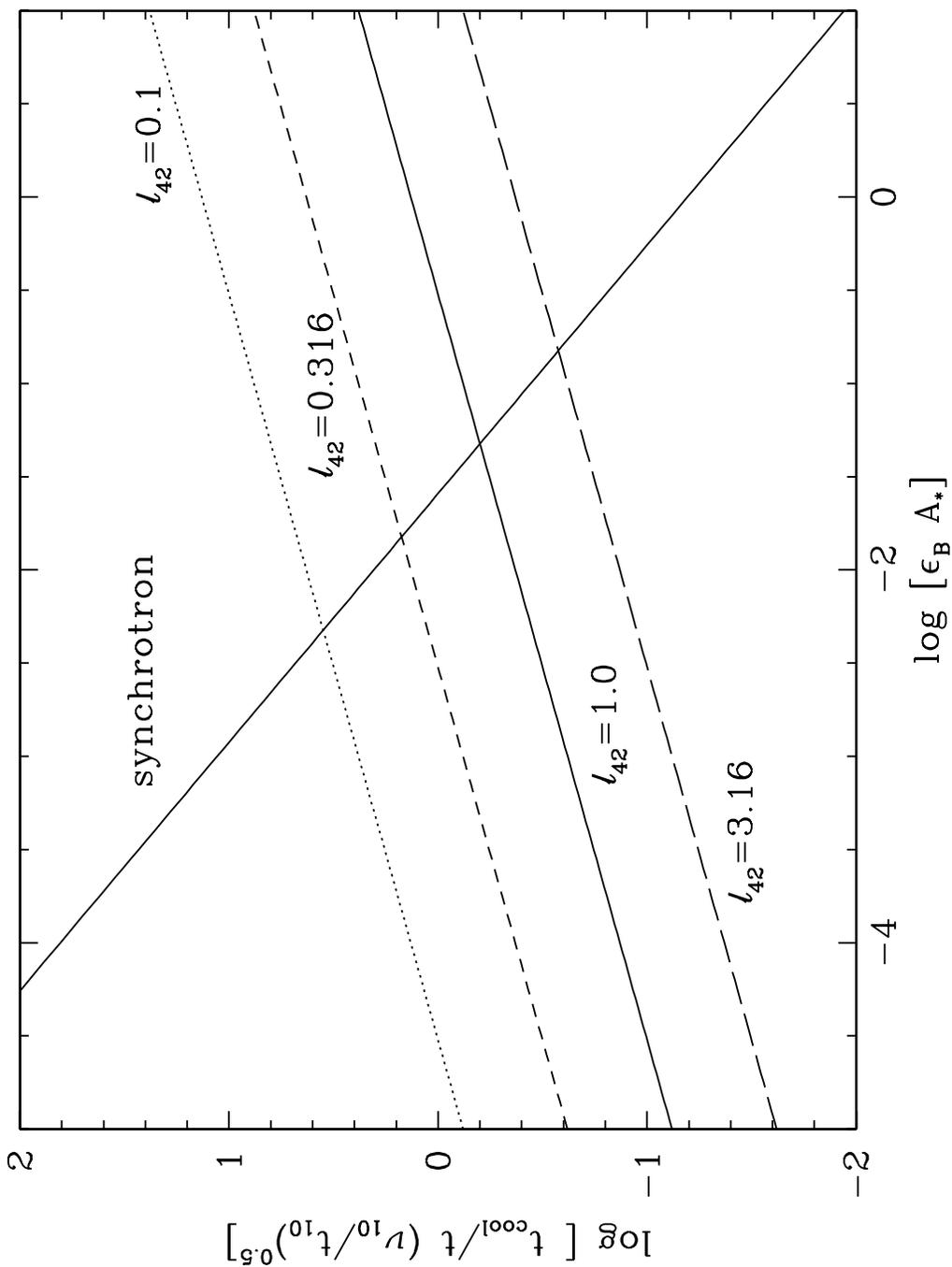} \figcaption{Ratio of cooling time scale to
adiabatic time scale for synchrotron cooling (single solid line)
 and for inverse Compton cooling (4 parallel lines). 
 For Compton cooling each curve is labeled with the value of
$l_{42} = (L_{\rm bol}/10^{42} {\ergs})(V_s/15,000 \kms)^{-2}$.
\label{figcomp}}
\end{figure}

\begin{figure}[!hbtp]
\includegraphics[scale=.70]{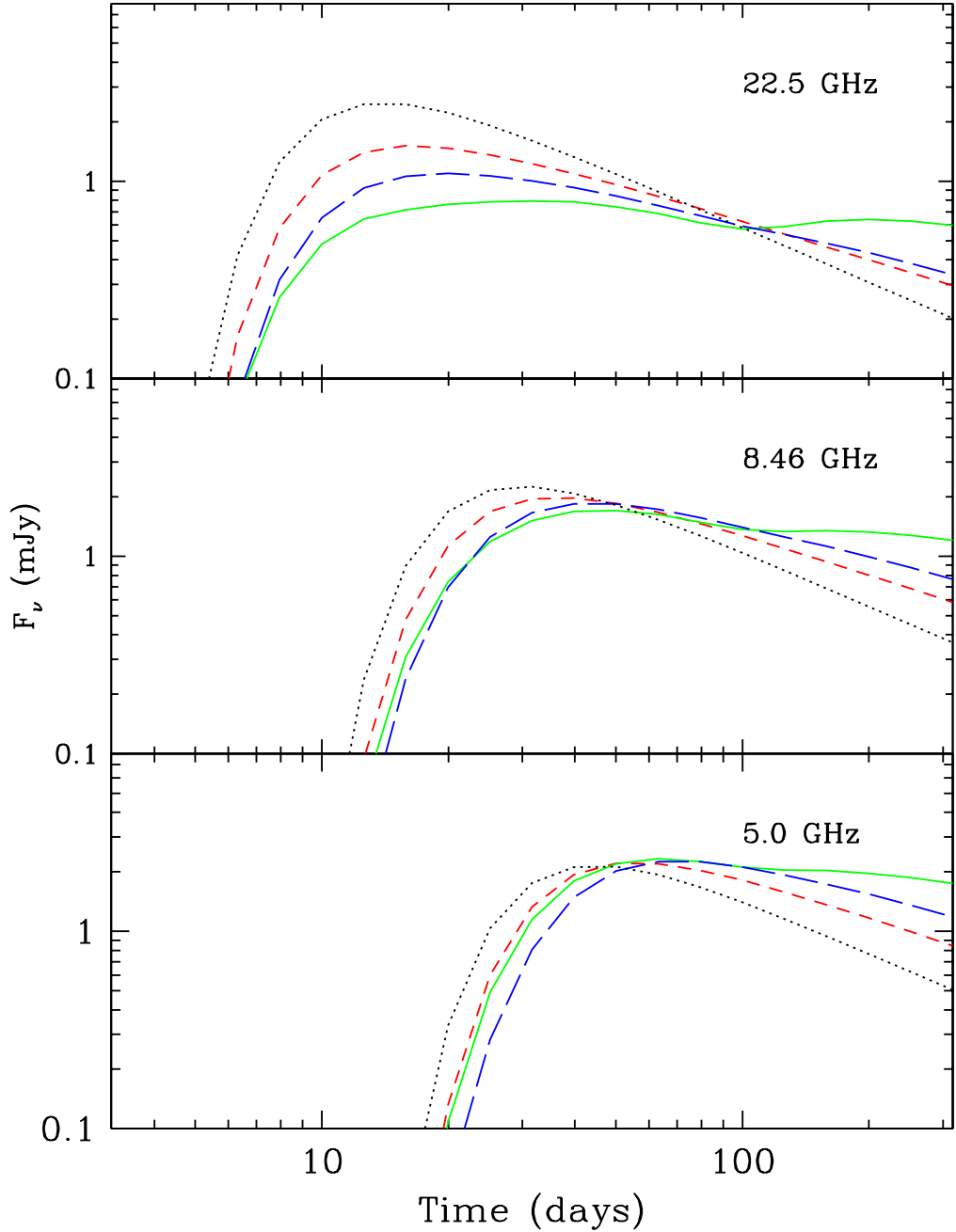} \figcaption{Radio light curves of four models
with different cooling properties. The dotted line corresponds to a
purely adiabatic model, the short dashed to a `minimal cooling' model,
the long-dashed curves to a model
dominated by synchrotron cooling, and the solid curves to a model
dominated by Compton cooling. See text for the parameters of each model.
\label{figlc}}
\end{figure}

\begin{figure}[!hbtp]
\includegraphics[scale=.70]{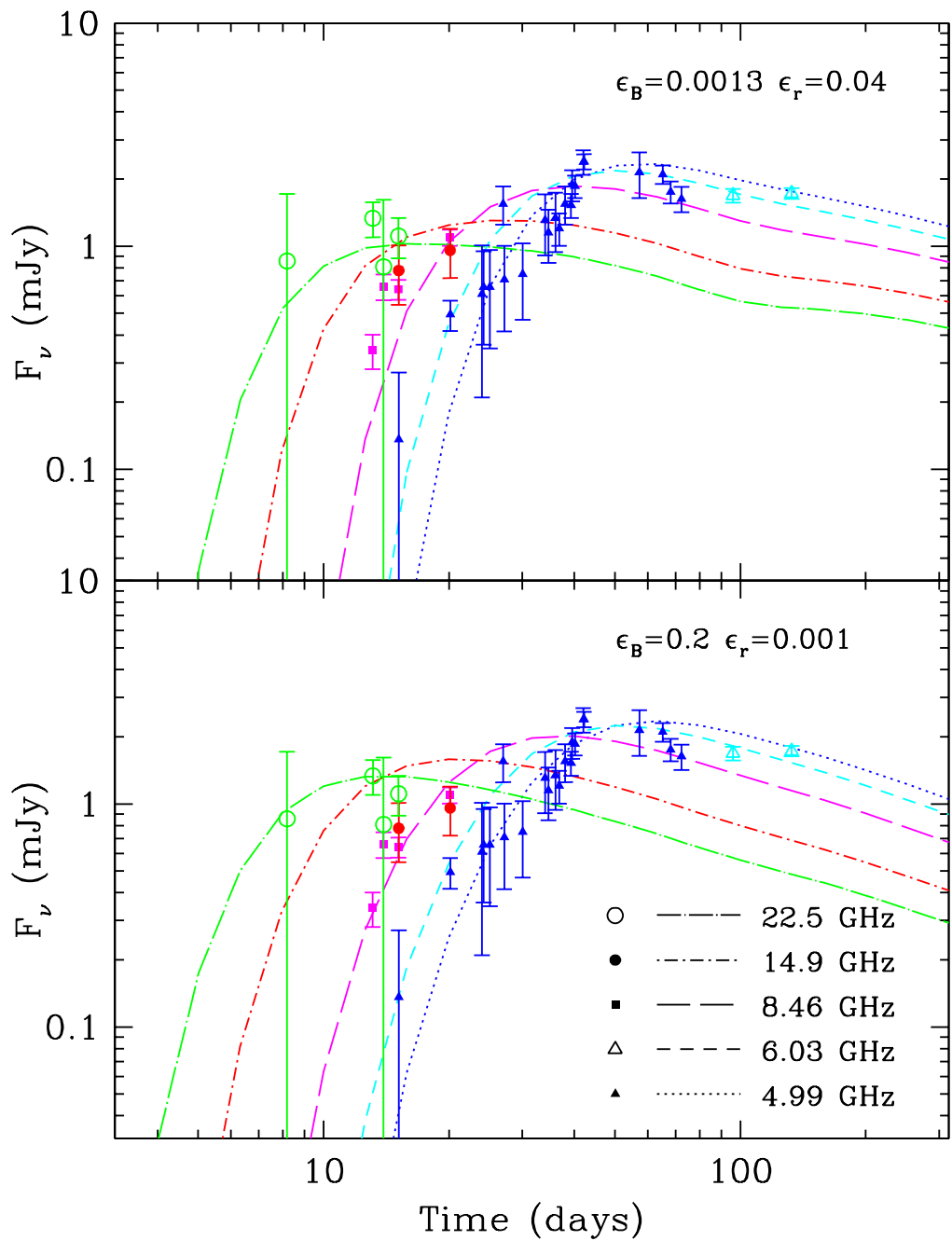} \figcaption{Observed
radio light curves for SN 2004et compared to models with two different
combinations of $\epsilon_B$ and $\epsilon_r$. Note especially the
effects of Compton cooling for the upper panel, resulting in nearly
flat optically thin light curves, as well as the dip at $\sim 100$
days seen at the high frequencies. 
\label{sn2004et_lc_fit}}
\end{figure}

\begin{figure}[!hbtp]
\includegraphics[scale=.70]{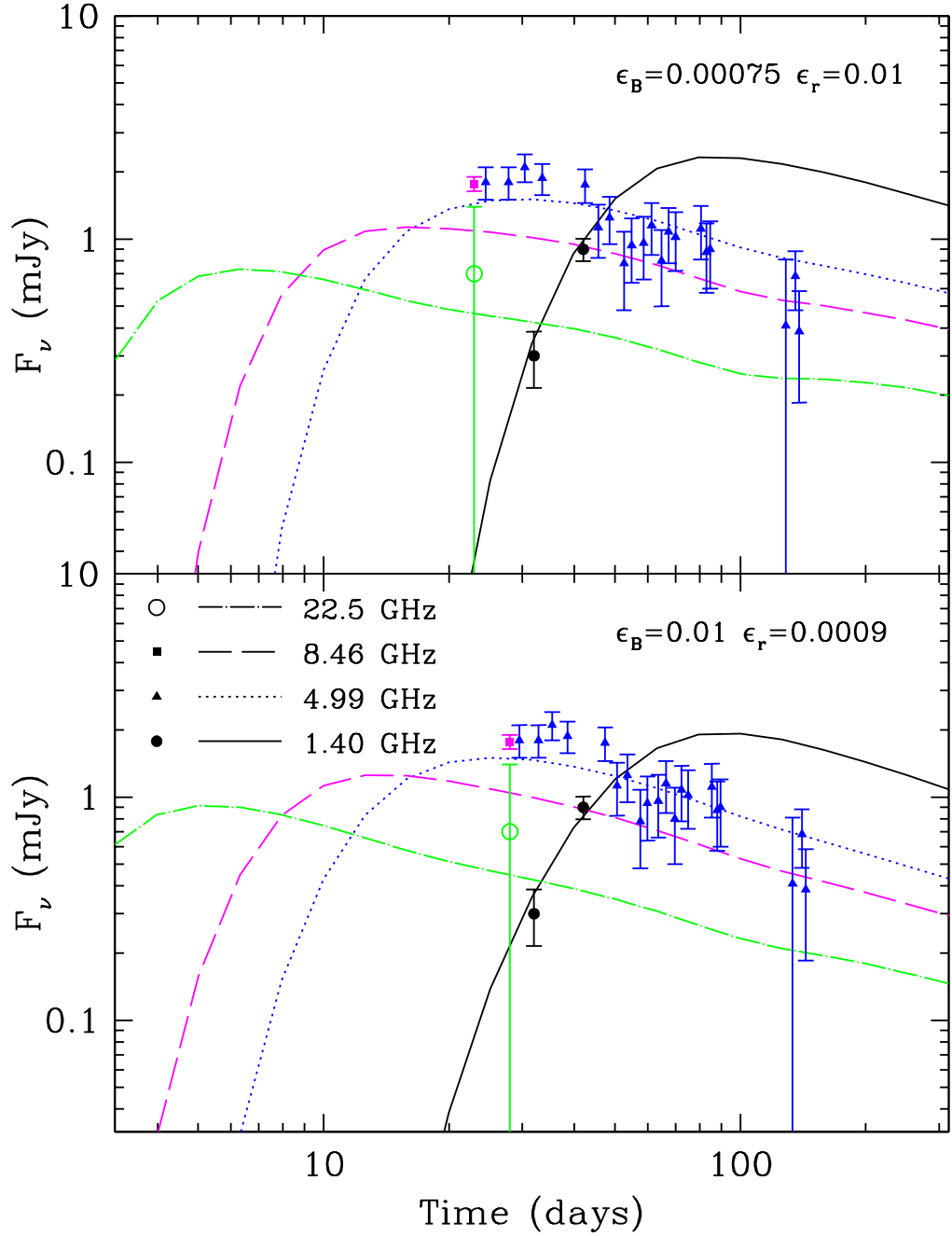} 
\figcaption{Observed
radio light curves for SN 2004dj compared to models with two different
combinations of $\epsilon_B$ and $\epsilon_r$.
\label{sn2004dj_lc_fit}}
\end{figure}

\begin{figure}[!htp]
\includegraphics[scale=.60,angle=270]{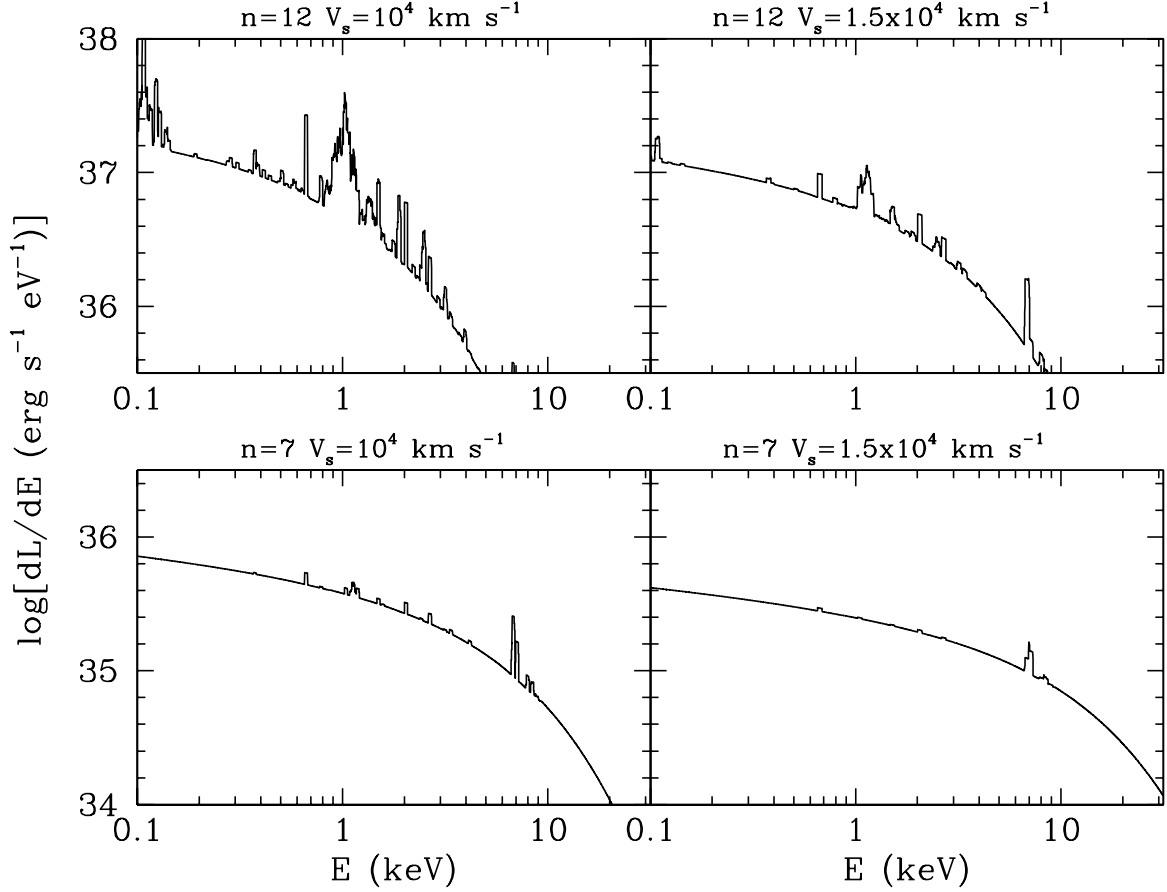} 
\figcaption{Calculated X-ray spectra for
$n=7$ and 12, and $V_s=10,000 \kms$ and $V_s=15,000 \kms$. The mass
loss rate is $5\times 10^{-6} \ml$ (for $v_{w} = 10\kms$). Note the
increasing importance of line emission as the reverse shock
temperature decreases. For the highest temperature ($n=7$ and
$V_s=15,000 \kms$) the only important lines are the Fe K lines at $\sim 7.0$
keV.
\label{figxray}}
\end{figure}

If the progenitor mass of SN 2003gd was $8-9\Msun$, the mass loss density was a factor
of $\sim 3$ lower than for a $15\Msun$ star (Fig. \ref{massloss}), 
leading to a radio luminosity $5-10$
times smaller at the same age.
This is consistent with the upper limit on the radio luminosity and
shows the difficulty of observing low mass SNe IIP at radio wavelengths.
The observations must be undertaken at an early time and with high sensitivity.
The recent Type IIP SN 2005cs in M51 is also estimated to have a relatively
low mass progenitor, $8\pm 1\Msun$ \citep{Li05b} or $9^{+3}_{-2}\Msun$ \citep{MSD05}.
It has not been reported as an X-ray or radio source.

The initial mass deduced for SN 1999em by \cite{Sma03}  of
$<15\Msun$ is lower than that found for SN 2004dj by \cite{Mai05},
but \cite{Leo03} found a mass $<20\Msun$ for SN 1999em.
The value of $\dot M_{-6}/v_{w1}$ (Table 2) is comparable to that
for the other SNe IIP and is consistent
with stellar evolution and typical red supergiant
mass loss rates for a $15\Msun$
initial mass star.

SN 1999gi was considerably less luminous in X-rays than SN 1999em (Table 1), implying
a lower value of $\dot M_{-6}/v_{w1}$ by a factor $\sim 3$
for comparable supernova properties.
Considering the relation between $M$ and $\dot M$ shown in Fig. \ref{massloss},
a drop in $\dot M$ by a factor 3 corresponds to going from a $15\Msun$
initial mass star to an $8\Msun$ initial mass star.
This is consistent with the limit $<12\Msun$ deduced for the progenitor
of SN 1999gi.

An independent way to estimate the mass of an exploded star is from
the properties of the supernova.  From analyses of the light curves of
the supernovae, ejecta masses for SN 1999em have been estimated as
$15.0\Msun$ \citep[][for $d=11.1$ Mpc]{N03}, $27^{+14}_{-8}\Msun$
\citep[][for $d=10.7$ Mpc]{H03} and, for SN 1999gi, $18.7\Msun$
\citep[][for $d=11.8$ Mpc]{N03}, $43^{+24}_{-14}\Msun$ \citep[][for
$d=9.0$ Mpc]{H03}.  With this method, the mass estimate decreases as
the distance estimate increases.  To obtain the initial mass, the mass
of any compact remnant and mass
lost during the evolution must be added to the ejecta mass.  It can be
seen that the masses obtained in this way are larger than the masses
obtained from the mass loss or from the direct progenitor
observations, although Nadyozhin's value for SN 1999em is fairly close
to the value deduced by these other methods. The method is, however,
sensitive to several factors.  In particular,
hydrodynamic mixing, as was found to be very important for the light
curve of SN 1987A, may change the mass estimates considerably 
\citep[e.g.,][]{Chi03}.  More complex supernova models, including models of
the spectra, are needed to see whether better agreement can be found
between these methods.  The current work makes specific predictions
for SN 1999em and SN 1999gi, which have only upper limits from the
direct progenitor observations.

Table 1 includes estimates of the metallicity of the regions in which the
supernovae occurred.
The metallicities are either solar or within a factor of 2
of solar;  this variation is expected
to have a minor effect on the present discussion if the estimated
$Z^{1/2}$ dependence of mass loss is correct.
If SNe IIP are discovered in highly metal poor regions, there is the
possibility of investigating the metallicity dependence of mass loss.

As the sample of supernovae grows, it will be possible to compare
the distribution of observed progenitor masses with that expected for
a standard initial mass function.
If we consider two mass groups, $8-13\Msun$ and $13-18\Msun$,
4 of the 7 supernovae mentioned here (SNe 1999em, 2002hh, 2004dj, and 2004et)
belong to the high mass group and 3 (SNe 1999gi, 2003gd, and SN 2005cs) belong
to the low mass group.
For a Salpeter mass function, there would be a factor 1.7 fewer
supernovae in the high mass group compared to the low mass group.
At least part of the reason for the observed relative paucity of low
mass progenitors may be that they give rise to subluminous supernovae.
SN 2005cs had a peak absolute $V$ magnitude of $-15.6$ \citep{Li05b},
which is subluminous (see \S~3).
Also, it is possible that not all stars in the $8-13\Msun$ range end their
lives as SNe IIP.
The lower mass cut for supernovae is not well-determined,
and stars in this mass range may end their lives as a type of
supernova other than a IIP.

In summary, data on the wind interaction and progenitor stars of
a small number of SNe IIP are consistent with mass loss estimates and
tracks in evolutionary models of $\sim 8-20\Msun$ single stars.
The radio and X-ray emission from SNe IIP is relatively faint,
but this emission may provide a check on the mass of the
supernova progenitor.
In addition, the fact that the radio emitting electrons are
likely to be subject to energy loss affects the radio light
curves and gives constraints on the physical properties of
the shock interaction.
X-ray emission gives additional diagnostics and will be most
useful if well-determined spectra can be obtained.

\acknowledgments
We are grateful to Megan Argo, Rob Beswick,
Chris Stockdale, Kurt Weiler, and their collaborators for
making radio data available in advance of full publication.
This research was supported in part by NSF grant AST-0307366,
Chandra grant TM4-5003X, and the
Swedish Research Council and National Space Board.

\clearpage

\vspace*{1 cm}
\noindent{Table 1. Type IIP supernovae }

\begin{tabular}{cccccc}
\hline
SN &   Distance  & Metallicity & $L_x$(0.5-8 keV) & Progenitor   & Refs.  \\
  &   (Mpc)  & ($Z/Z_{\odot}$)& (ergs s$^{-1}$) & mass ($\msun$)     &      \\
\hline
1999em  &$11.7\pm1.0$ & $1-2$  &  $9\times 10^{37}$   & $< 15$    &  1,2,3   \\
1999gi  & 11.1  &  $\sim 2$  & $1.6\times 10^{37}$   &  $< 12$     &  2,4,5   \\
2002hh  & $5.5\pm 1.0$  &    &  $4\times 10^{38}$   & --     & 6,7   \\
2003gd  &  9 & $\sim 0.5$  &   --   & $8-9$     & 8,9  \\
2004dj  &  3.2 &  1  & $1.5\times 10^{38}$   &  $15,\sim 12$  & 10,11,12  \\
2004et  & $5.5\pm 1.0$ &    &  --   &  $15^{+5}_{-2}$    & 6   \\
\hline

\end{tabular}

References: (1) Leonard et al. 2003;   
(2) Smartt et al. 2003;  
(3) Pooley et al. 2002;  
(4)  Leonard et al. 2002;  
(5) Schlegel 2001;  
(6) Li et al. 2005;  
(7)  Pooley \& Lewin 2002;  
(8) Van Dyk et al. 2003;  
(9) Smartt et al. 2004;   
(10)  Ma\'iz-Appel\'aniz et al. 2004;  
(11)  Wang et al. 2005;  
(12) Pooley \& Lewin 2004.  


\vspace*{1 cm}
\noindent{Table 2. Results for radio supernovae }

\begin{tabular}{ccc}
\hline
SN &   $t(\tau_{ff}=1)$ at 1.4 GHz   & $\dot M_{-6}v_{w1}^{-1}T_{cs5}^{-3/4}$  \\
  &   (day)  &    \\
\hline
1999em &52 & 5       \\
2002hh &62 &  7     \\
2004dj  & &  $2-3$      \\
2004et &  &   $9-10$     \\
\hline

\end{tabular}

\clearpage

\end{document}